\documentclass[doublecol,linenumbers]{epl2} 

\title{Double valley Dirac fermions for 3D and 2D Hg$_{1-x}$Cd$_x$Te \\with strong asymmetry}
\shorttitle{Double valley Dirac fermions for 3D and 2D Hg$_{1-x}$Cd$_x$Te with strong asymmetry} 

\author{M. Marchewka}
\institute{                    
Faculty of Mathematics and Natural Sciences,
Centre for Microelectronics and Nanotechnology, \\
University of Rzesz\'ow, Pigonia 1, 35-959 Rzesz\'ow, Poland.\\
}

\pacs{73.20.-r}{Electron states at surfaces and interfaces}
\pacs{73.61.Ga}{II-VI semiconductors}
\pacs{74.20.Pq}{Electronic structure calculations}

\abstract{
In this paper the possibility to bring about the double- valley Dirac fermions in some quantum structures is predicted. 
These quantum structures are: strained 3D Hg$_{1-x}$Cd$_x$Te topological insulator (TI) with strong 
interface inversion asymmetry and the asymmetric 
Hg$_{1-x}$Cd$_x$Te double quantum wells (DQW). 
The numerical analysis of the dispersion relation for 3D TI
Hg$_{1-x}$Cd$_x$Te for the proper Cd ($x$)-content of  in the Hg$_{1-x}$Cd$_x$Te-compound  
clearly show that the inversion symmetry breaking together with the unaxial tensile strain 
causes splitting each of Dirac nodes (two belonging to two interfaces) 
into two in the proximity of  $\Gamma$-point. 
The similar effects can be obtained for asymmetric Hg$_{1-x}$Cd$_x$Te DQW with the 
proper content of Cd and proper width of the quantum wells. 
The aim of this work is to explore the inversion symmetry breaking in 3D TI and 2D DQWs mixed HgCdTe-systems. It is shown that this symmetry breaking leads to the dependence of carriers energy vs quasi-momentum similar to that of Weyl fermions.}

\begin{document}

\maketitle

\section{Introduction }

 In Ref.\cite{Fu} the existence of 3D Topological Insulators (TI) in the HgTe 75-80 nm wide strained layers has been  predicted theoretically, and it was later confirmed experimentally in the work \cite{Brune}. Since then many papers devoted to the the topological 
surface states (TSS) existing at the surfaces of such structures and which can be described by the 
pseudo-Dirac fermions\cite{Brune,BruneRevX,Fr,Kozlov,Ros3D}  were published. 
The Dirac-like dispersion is observed due the uniaxial tensile strain along (001) axis which lifts 
the degeneracy of the $\Gamma_8$  band by breaking the cubic symmetry at the $\Gamma$ point  and opens the insulating gap. 
Against this gap the TSS are visible in a small energy area  about 22 meV (for 3D HgTe TI) that makes such systems 3D TI\cite{Brune}. 

The Dirac cone corresponds to a Dirac semimetal because there is no gap between the two cones, which would become hyperbolae when a gap is present. A normal insulator has a gap and a three dimensional topological insulator is
characterized by the bulk of the material having a gap while the surface does not. A Dirac semimetal, such as for example, Na$_{3}$Bi, is a three dimensional system with a Dirac cone having a double degeneracy at the Fermi energy;  Weyl semimetal has its valence and conduction bands touching each other at isolated points, around which the band structure
forms non-degenerate three dimensional Dirac cones. The apexes of the Dirac cones are called Weyl nodes.

In case of the 2D systems based on CdTe/HgTe, the sizes of the HgTe QW  makes the Dirac-like dispersion possible to be observed
 (for about 6.4 nm wide HgTe QW), which has been first predicted theoretically\cite{Bernevig} and after that was proven by the experiment\cite{Koenig}.

The Dirac cones are described by the four-component Dirac spinors satisfying
the Dirac  equation; when the mass is set equal to zero in the Dirac equation, it decouples into two equations
known as the Weyl equations that have two component spinors as solutions.
 
In order to pass from Dirac semimetal to Weyl semimetal, the breaking of the time-reversal symmetry (TRS) or the breaking of the inversion symmetry is required.  
The breaking of TRS can be achieved in HgCdTe by magnetic impurities doping, namely in Hg$_{1-x-y}$Cd$_x$Mn$_y$Te\cite{Weyl}.
On the other hand, in order to achieve this effect in the superlattice based on HgTe/CdTe multilayer structure, the broken inversion symmetry 
is required\cite{Pankratov,Babor}.

The  breaking of inversion symmetry enforces each Dirac node to split into two pairs of separate Weyl  nodes of same chirality at opposite momenta +/- {\bf k}$_0$.
This comes form the fact that if a Weyl node occurs at some momentum {\bf k}, time-reversal symmetry requires that another Weyl node occurs at {\bf -k} with equal topological charge. It holds that the total topological charge associated with the entire Fermi surface must vanish. Hence, there must exist two more Weyl points of opposite topological charge at {\bf k$_0$} and $-${\bf k}$_0$. 

 In this paper the double-valley Dirac Fermions in the strained
3D Hg$_{1-x}$Cd$_x$Te topological insulator with interface
inversion asymmetry as well as for the asymmetric Hg$_{1-x}$Cd$_x$Te DQWs  
are considered.

It was shown in author's  previous work  that  for symmetric structures with the same interfaces and
the proper content of $x$-Cd in  Hg$_{1-x}$Cd$_x$Te-compound  together with the uniaxial tensile strain along (001) direction and the 
proper width of the strained layers are enough to obtain the Dirac cone inside  the gap between $\Gamma_8^{lh}$ and $\Gamma_8^{hh}$\cite{Mar}.
In this papers it was also clearly shown that to obtain the interplay between two Dirac-like dispersions, such condition is extremely important and without that the double-valley Dirac semimetal (Weyl semimetal) can not be realized. Here it is necessary to add some comments concerning terminology which is used here. In the literature devoted to the subject of present discussion, the term Weyl semimetal is commonly accepted for describing the situation when the Dirac cone is split into two under the symmetry breaking. One should remember however, that Weyl equations are valid for massless particles; in case of  Hg$_{1-x}$Cd$_x$Te-based structures the masses of carriers, whatever small, are finite and it corresponds to the finite energy gap in TI. That is why we use the term 'Weyl semmetal'  with caution, sometimes substituting it by double-valley Dirac semimetal.

So far, for the QW based on the HgTe the authors (e.g\cite{Koenig}) focused on Dirac-like 
dispersion generally  for about 6.4 nm wide QW or for wider QW
with inversion dispersion relation (see, for example Refs \cite{Portal,Ros2D1, Ros2D}). The last one concerns
although the HgCdTe based quantum wells near the phase transition
between the semi-metal and semiconductor\cite{Wen}. Last time the influence 
of the natural interface inversion asymmetry (IIA) of zinc-blende  heteorostructures which drastically modifies 
both bulk and edge states are also investigated\cite{Tarasenko15,Tarasenko16}.
Besides the IIA, the contribution to so-colled ``zero magnetic field'' spin splitting 
in 2D electron system can be obtained by the bulk inversion asymmetry (BIA)
and structural inversion asymmetry (SIA)\cite{Winkler}. In case of the BIA, many authors suggested that it does not play an important role in the 2D zinc-blende materials\cite{BIA1}. In contrast to BIA, SIA known as Rashba
spin-orbit (SO) splitting, usually dominates in two-dimensional structures with inversion layers or asymmetric QWs
with an asymmetric confinement potential \cite{Bychkov}. It is responsible for most important effect which can be used in spintronics \cite{Rashba} and usually is achived (in case of the II-VI heteorostructures) by means of various delta-doping or applied gate voltage \cite{Hinz, Schultz}. 

So far in case of the DQW based on HgCdTe, theoretical as well as experimental papers were focused on two identical QWs\cite{Yakunin1, Yakunin2} or two HgTe QWs with the smaller as well as greater widths than its critical value\cite{Elena1}; however, the more complicated systems were also studied  (see, for instance, Ref.\cite{Pfeffer}). 

The discussion carried on in most of the papers devoted to 3D HgCdTe TI\cite{Brune,BruneRevX} as well as to 2D HgTe TI\cite{Koenig} were based on two complementary models: 1. the {\bf kp}-model (which allows to 
get the Dirac point for 6.4 nm wide QW and the TSS for 3D TI against strained gap);
2. simple one/two Dirac cones model for QW/3D TI structures respectively. These allow to explain  the  magneto transport experiments\cite{Koenig, Brune, SolidMar, wrobel}.

In this work the eight-band {\bf kp} model\cite{Novik, APJ} is used to calculate the dispersion relation  for the two different types of HgCdTe-based TI: 1. the strained 3D TI 100 nm wide HgTe as well as the mixed HgCdTe layers with two different interfaces; 2. for DQWs with the different $x$-Cd content in the QWs and different widths of them. 

The paper deals with the HgCdTe-structures with broken inversion symmetry, but not the broken time-reversal symmetry as in other papers. 
All results presented here relate to  4.2 K, and were obtained by means of the methods elaborated in Ref. \cite{MarIJMP}.
Two structures which were considered, fulfill the definition of topological  insulator-normal insulator (TI-NI) multilayer structures\cite{Zyuzin}. 

For the case of 3D TI, two different interfaces correspond to IIA, while for the second, asymmetric DQWs structures, for which 
the chosen geometry corresponds to SIA. Such band-engineering allows to get  two linear dispersion cones in the nearest proximity of $\Gamma$-point in the Brillouin zone.

Such linear-dispersion together with band-touching points called Weyl nodes are very promisning from the point of view of future applications. The charge carrier properties of such system  are completely different in comparison with known semiconductor structures used untill now \cite{Ominato}. 

\section{3D Hg$_{1-x}$Cd$_x$Te TI }

As it was mentioned above, so far many  publications were devoted to the symmetric 
strained layers (about 75-80 nm wide HgTe) placed between two identical materials.
The authors of these papers restricted their consideration only to different potentials originated from the different dielectric constancts $\epsilon$ on both surfaces of the structure in question, or to the external gate voltage (see, for instance, Ref. \cite{BruneRevX}).

\begin{figure*}
\begin{center}
\includegraphics[scale=0.55]{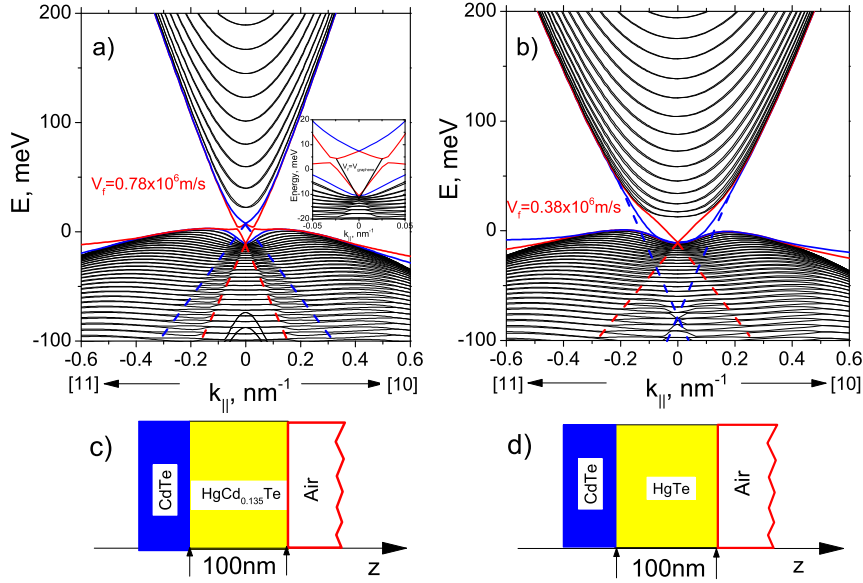}
\caption{E(k) for: a) strained 100 nm wide Hg$_{0.865}$Cd$_{0.135}$Te and b) strained 100 nm wide HgTe; 
obtained for non-symmetric cases structures -  the sketch are visible in c) and d). The blue and the red bold curves represent the 
TSS for the interfaces between the CdTe and vacuum, respectively. The dotted colored curves show the lower part of the
Dirac cones which is localized: in the case of a) - on the top of the $\Gamma^{hh}$ (red curve) and inside the strained gap - blue curve;
and in the case of b) - on the top of the $\Gamma^{hh}$ (red curve) and deep inside the $\Gamma^{hh}$ band - at $\approx$ -80 meV.}
\label{f1.lbl}
\end{center}
\end{figure*}

Without external voltage the different 
interfaces manifest themelves in the vicinity of the $\Gamma_{hh}$ where the Dirac point is located at the
$\Gamma$-point before hybridization\cite{Brune}. 

The influence of the external voltage applied to 3D HgTe is observed  only for TSS in {\bf k}-space far away from $\Gamma$ point, when 
the Fermi level is located between the $\Gamma_6$-  and $\Gamma_8$-states. In this case the 
external gate voltage is necessary for two reasons: first, to decrease the concentration,  which is especially important for low carrier densities when the broadening is stronger due to reduced screening\cite{BruneRevX}; and second, to split the TSS in order to observe two filling factors originating from  two different 
independent 2DEG at the surfaces corresponding to two Dirac-cones under magnetic field according to the relation $\nu=(N_t+N_b+1)$.

Similar approach was used in case of pure 3D strained HgTe structures with the 
Nb as superconductor\cite{Maier}.  Here however, the low transparency 
the Nb-HgTe interfaces is responsible for the observed periodic oscillations of differential resistance. 

First of all, the mixed Hg$_{1-x}$Cd$_x$Te strained 100 nm wide layers with the symmetric interfaces 
and $x$-Cd content equals to 0.13 allows to get the Dirac point inside of the strained gap\cite{Mar}.
In this case we have two single no crossing Dirac-valleys at the $\Gamma$-point. 
In the similar structure, but with different materials at the two interfaces,  the  inversion symmetry breaking could manifest itself in more spectacular manner.
The results of numerical calculation of the energy  dispersion for such case  are shown in Fig. \ref{f1.lbl} for: a) 
the strained mixed Hg$_{0.865}$Cd$_{0.135}$Te layers and b) strained 100 nm wide pure HgTe layers. 
In Figs. \ref{f1.lbl}c) and \ref{f1.lbl}d)  the sketches of the structures are shown for these two cases.
The vacuum should be treated as a second different interfaces. It was defined by means of boundary condition for {\bf kp} method\cite{boundary}.

For pure HgTe (Fig. \ref{f1.lbl}b) the different shapes of Dirc-cones are clearly visible; they correspond to two different  
surfaces. The TSS belonging to CdTe/HgTe interface are visible at the strained gap; they originate from Dirac point
which is deep inside the $\Gamma_8^{hh}$-band, while the TSS at the HgTe/vacuum interface starts from the Dirac point 
which in this case is localized at the top of the $\Gamma_8^{hh}$-band  in the center of the  Brillouin zone. 

This situation becomes more complicated  for the 100 nm wide strained Hg$_{0.865}$Cd$_{0.135}$Te layers - see Fig. \ref{f1.lbl}a). 
The $x$-Cd content equal to $x=$0.135 raises the Dirac point in energy scale. 
In addition, two different interfaces lead to the splitting of  Dirac-cone into two (they are depicted as red and blue lines in Fig. \ref{f1.lbl}a). The apexes of the cones 
correspond to different energies in $\Gamma$-point  and it is important that $E(k)$-curves 
are characterized by different slopes  in the vicinity of the $\Gamma$-point.

\begin{figure*}
\begin{center}
\includegraphics[scale=0.36]{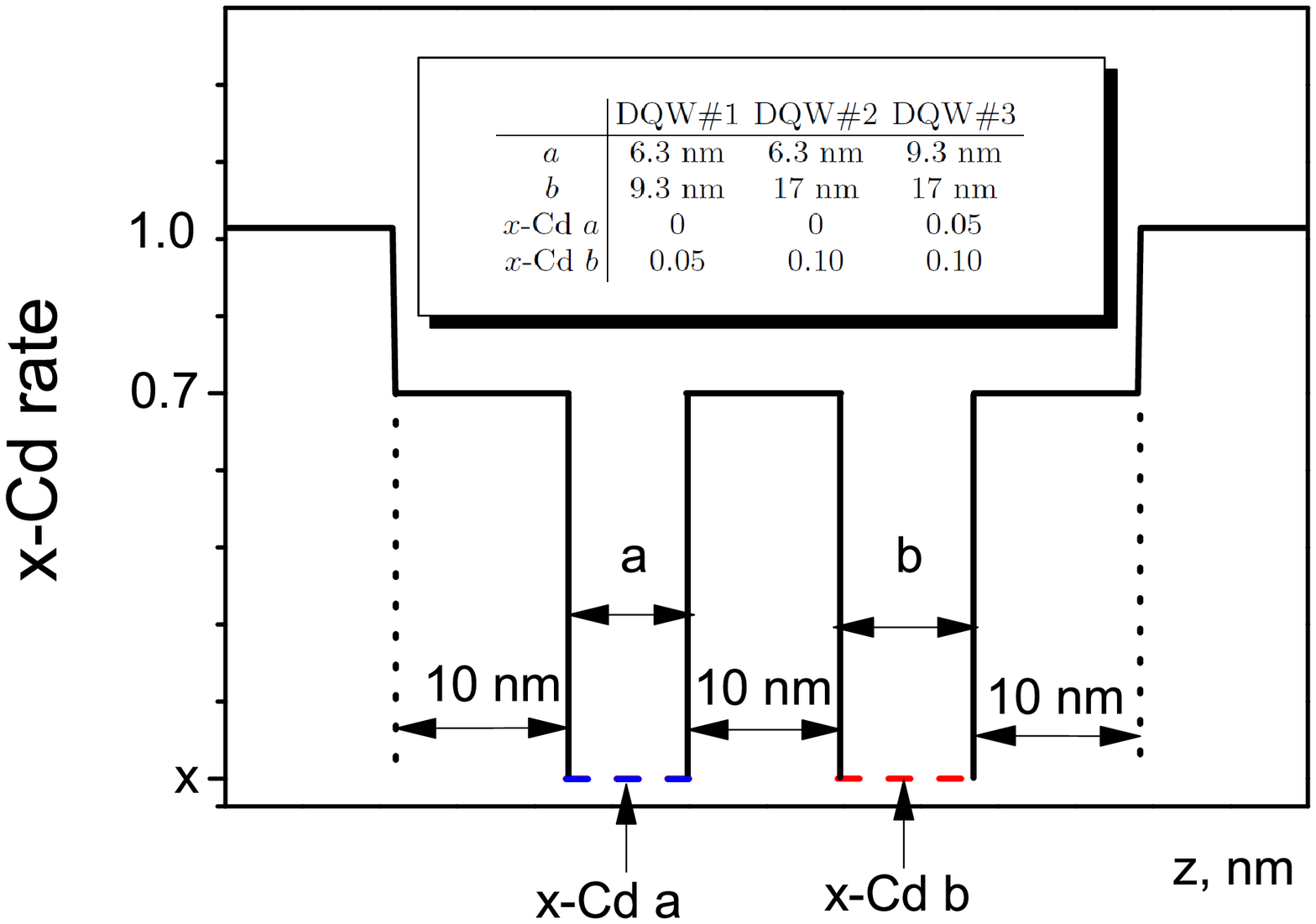}
\includegraphics[scale=0.38]{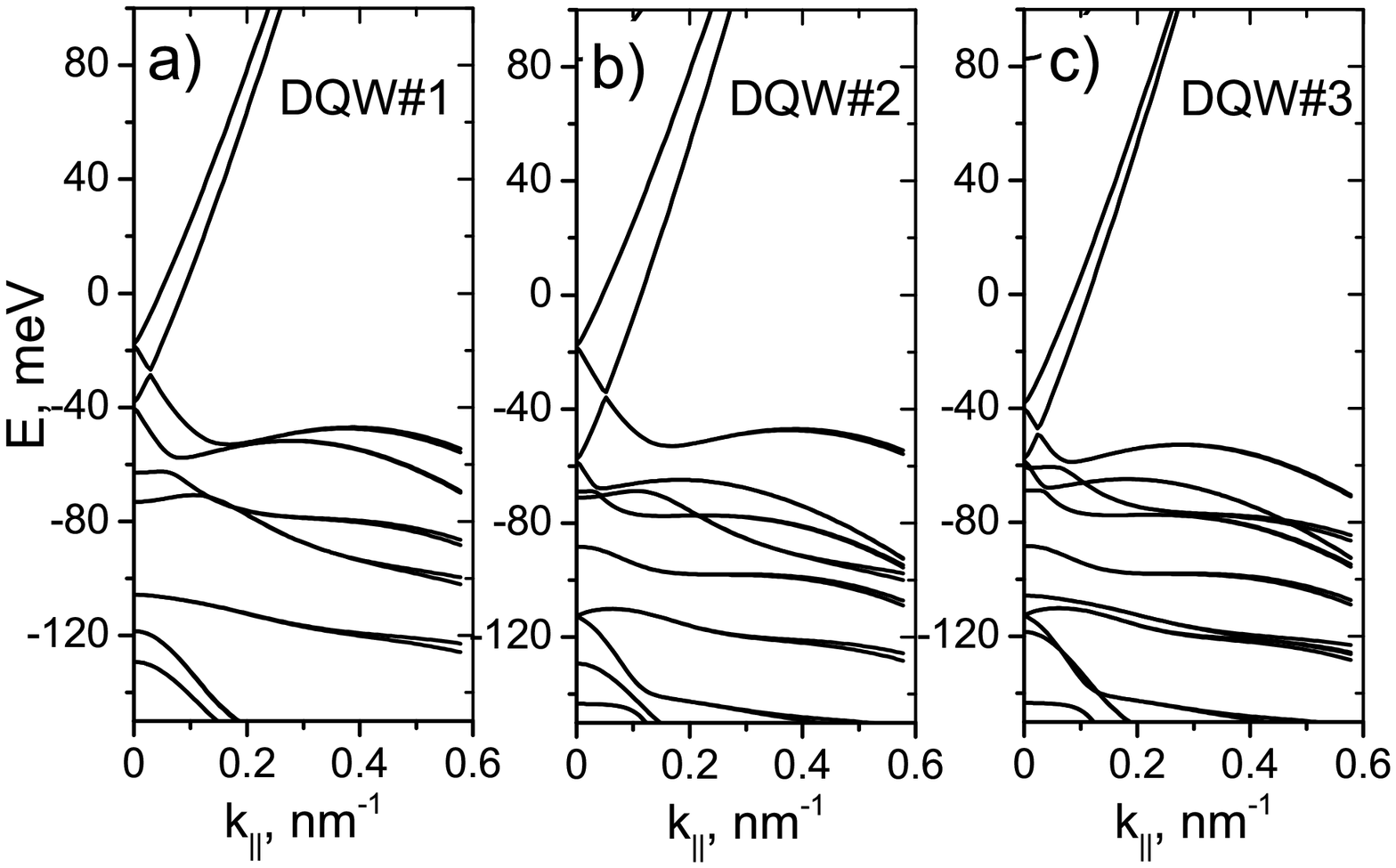}
\caption{The scheme of the DQWs structures with the values of the x-Cd compounds and width of the QWs as well as 
the barriers which were taking into account into calculations (left panel). Right panel: E(k) for DQWs for three cases: a) 
DQW\#1; b) DQW\#2 c) DQW\#3 - obtained for the parameters summarized in the table  - left panel.}
\label{f2.lbl}
\end{center}
\end{figure*}

This differences  enforces different dispersion curves to intersect each other, so we can observe  two 
Dirac valley nodes symmetrically arranged in a closed proximity of $\Gamma$-point.
It is worth to mention that the TSS characterizing the interface at the vacuum side of the structure can be fitted by the Dirac-like 
Hamiltonian with the Fermi velocity two times 
greater  than  that for the pure 3D HgTe TI (see Figs. \ref{f1.lbl}, a) and b). This velocity 
can be calculated directly using the relation $\Delta E=\hbar$v$_f\Delta k$ 
in the domain where dispersion is linear with respect to {\bf k}. These results are shown in Fig. \ref{f1.lbl}  
for both cases; the values written in red (Fig. \ref{f1.lbl}a-b) mean  the Fermi velocity for interface on the vacuum side. 
The black curve ( see panel inserted  in Fig. \ref{f1.lbl}a) corresponds to the 
linear dispersion for which Fermi velocity is equal 1.06 x 10$^6$ m/s.

\section{\label{sec:level3} Hg$_{1-x}$Cd$_x$Te DQWs}

As it was mentioned above, so far the DQWs with the HgTe/HgCdTe materials were investigated for
the structures with pure HgTe in QWs and different widths. 

 Just like in previous case of the 3D TI, it should be interesting to explore the possibility to get two different Dirac-cones in case of quantum 2D system.
 
Based on the results presented in author's paper (Ref.\cite{MarIJMP}), in present paper such quantum 2D system, namely DQWs were designed in order to search for
the Dirac-like dispersion.  This one occurs due to different  $x$-Cd content in the Hg$_{1-x}$Cd$_x$Te-compound and different widths of QWs.   
The sketches of the  DQWs structures for which the calculation were done, are presented in Fig. \ref{f2.lbl} (left part). The parameters of the CdTe/Hg$_{1-x}$Cd$_x$Te  DQWs structure, the widths of QWs and the $x$-Cd content are summarized in the upper left panel inserted in Fig. \ref{f2.lbl}. The calculations of the dispersion relations were
carried out using these parameters as well as {\bf kp} band parameters\cite{SolidMar}; the results are  presented in Figs. \ref{f2.lbl} a)-c).
Using the obtained linear dispersion for the small values of {\bf k}-vector (ranging from 0 up to 0.1 nm$^{-1}$), we conclude that Fermi velocity is equal 0.95 x 10$^{6}$ m/s. This values is similar to that one for graphene (1.06 x 10$^{6}$ m/s).  To verify experimentally the existence of two double valley Dirac cones  in case of the DQWs structure  should be easier, because of the simplicity of experimental setup.  Indeed, the contemporary  technology of producing and preparatory of single QW based on Hg$_{1-x}$Cd$_x$Te should be sufficient to observe the corresponding effects in DQW structures. 

\section{Summary}

The aim of this work is to explore  the consequences of inversion symmetry breaking in the 3D and 2D mixed HgCdTe systems.
As it is shown, it is possible to obtain the double valley Dirac cones in 3D TI as well as 
in DQWs based on HgCdTe. The necessary conditions to get such shapes of the dispersion relations are: for the 3D TI it is a proper $x$-Cd content in the  strained 3D HgCdTe layers together with the asymmetric interfaces; for HgCdTe DQWs it is a proper values of the x-Cd content  in each QWs and a proper values of the width of them.

So,  the touching points of the valence and conduction bands with linear dispersion relation in their vicinities can be found in mixed 3D HgCdTe TI as well as in HgCdTe DQWs systems. The effects related to the Dirac cones crossing in the structures examined in the paper, could be experimentally verified in the cyclotron resonance and/or magneto-transport experiments.

Such  band-touching points characterized by liner dispersion, called Weyl nodes were  for the first time predicted perhaps in Ref.\cite{ABRIKOSOV1970} and it looks like that they were already observed in strained Hg$_{1-x-y}$Cd$_x$Mn$_y$Te\cite{Weyl}.

Experimental observation of our theoretical predictions, 
would be an important advancement in understanding the nature of correlated Dirac fermions in
the unique state of matter, that is TI and Weyl semimetals.
The concequences of Weyl dispersion was recently analyzed by A.A. Burkov\cite{Burkov}, who predicted the existence of negative quadratic longitudinal magnetoresistance in the similar structures.  It seems, that the results of present paper are also could be interesting in regard of rapid development of such exciting area  of research as Topological Insulators and Weyl semimetals.


\begin{thebibliography}{0}

\bibitem{Fu} Liang Fu, C. L. Kane, E. J. Mele, Phys. Rev. Lett. {\bf 98}, 106803 (2007).
\bibitem {Brune} C. Br\"une, C. X. Liu, E. G. Novik, E. M. Hankiewicz, H. Buhmann, 
Y. L. Chen, X. L. Qi, Z. X. Shen, S. C. Zhang, and L.W. Molenkamp, Phys. Rev. Lett. 106, 126803 (2011) supp. material.
\bibitem {BruneRevX} C. Br\"une, C. Thienel, M. Stuiber, J. B\"ottcher, H. Buhmann, 
E. G. Novik, Chao-Xing Liu, E. M. Hankiewicz, L. W. Molenkamp, Phys. Rev. X {\bf 4}, 041045 (2014).
\bibitem {Fr} C. Bouvier, T. Meunier, P. Ballet, X. Baudry, R. Kramer, L. L\'evy, arXiv:1112.2092, (2011).
\bibitem {Kozlov} D. A. Kozlov, Z. D. Kvon, E. B. Olshanetsky, N. N. Mikhailov, S. A. Dvoretsky, and D. Weiss, Phys. Rev. Lett. {\bf 112}, 196801 (2014).
\bibitem {Ros3D} K. M. Dantscher, D. A. Kozlov, P. Olbrich, C. Zoth, P. Faltermeier, M. Lindner, G. V. Budkin, S. A. Tarasenko, V. V. Belkov, Z. D. Kvon, N. N. Mikhailov, 
S. A. Dvoretsky, D. Weiss, B. Jenichen, and S. D. Ganichev, Phys. Rev. B {\bf 92}, 165314 (2015).
\bibitem {Bernevig} B. A. Bernevig and Shou-Cheng Zhang, Phys. Rev. Lett. {\bf 96}, 106802 (2006).
\bibitem {Koenig} M. K\"onig, S. Wiedmann, C. Brüne, A. Roth, H. Buhmann, L. W. Molenkamp, Xiao-Liang Qi, Shou-Cheng Zhang, Science {\bf 318}, 5851 (2007).
\bibitem {Weyl} Daniel Bulmash, Chao-Xing Liu, and Xiao-Liang Qi, Phys. Rev. B 89, 081106(R) (2014).
\bibitem {Pankratov} O. A. Pankratov, B. A. Volkov,  ``Landau Level Spectroscopy'', {\it Part III: Two-Dimensional Systems}
 Eds G. Landwehr and E. I. Raschba, Amsterdam: North-Holland, Ch. 14, 818 (1991).
\bibitem {Babor} G\'abor B. Hala\'asz, Leon Balents, Phys. Rev. B {\bf 85}, 035103 (2012).
\bibitem {Mar} Micha\l{} Marchewka, Physica E {\bf 84}, 407 (2016).
\bibitem {Minkov} G. M. Minkov, A. V. Germanenko, O. E. Rut, A. A. Sherstobitov, M. O. Nestoklon, S. A. Dvoretski, and N. N. Mikhailov, Phys. Rev. B {\bf 93}, 155304 (2016).
\bibitem {Portal} O. E. Raichev, G. M. Gusev, E. B. Olshanetsky, Z. D. Kvon, N. N. Mikhailov, S. A. Dvoretsky, and J. C. Portal, Phys. Rev. B {\bf 86}, 155320 (2012).
\bibitem {Ros2D1} G. M. Minkov, A. V. Germanenko, O. E. Rut, A. A. Sherstobitov, S. A. Dvoretski, and N. N. Mikhailov, Phys. Rev. B {\bf 88}, 155306 (2013).
\bibitem {Ros2D} E. B. Olshanetsky, Z. D. Kvon, G. M. Gusev, A. D. Levin, O. E. Raichev, N. N. Mikhailov, and S. A. Dvoretsky, Phys. Rev. Lett. {\bf 114}, 126802 (2015).
\bibitem {Wen} Wen Yang, Kai Chang, and Shou-Cheng Zhang, Phys. Rev. Lett. {\bf 100}, 056602 (2008).
\bibitem {Tarasenko15} S. A. Tarasenko, M. V. Durnev, M. O. Nestoklon, E. L. Ivchenko, Jun-Wei Luo, and Alex Zunger, Phys. Rev. B {\bf 91}, 081302(R) (2015).
\bibitem {Tarasenko16} M. V. Durnev and S. A. Tarasenko, Phys. Rev. B {\bf 93}, 075434 (2016).
\bibitem {Winkler} R. Winkler, {\it Spin-Orbit Coupling Effects in Two-Dimensional Electron and Hole Systems}, Springer-Verlag, Berlin-Heidelberg (2003).
\bibitem {BIA1} M. Pang and X. G. Wu, Phys. Rev. B {\bf 88}, 235309 (2013).
\bibitem {Bychkov} Y. A. Bychkov and E.I. Rashba, J. Phys. C {\bf 17}, 6039 (1992).
\bibitem {Rashba} A. Manchon, H. C. Koo, J. Nitta, S. M. Frolov and R. A. Duine, Nature Materials {\bf 14}, 871 (2015).
\bibitem {Hinz} J. Hinz, H. Buhmann, M. Schoafer, V. Hock, C. R. Becker and L. W. Molenkamp, Semicond. Sci. Technol. {\bf 21}, 501–506 (2006).
\bibitem {Schultz} M. Schultzdag, F. Heinrichsdag, U. Merktdag, T. Colinddag, T. Skauliddag and S. Lovoldddag, Semiconductor Science and Technology {\bf 11}, 1168 (1999).
\bibitem {Yakunin1} M. V. Yakunin, A. V. Suslov, M. R. Popov, E. G. Novik, S. A. Dvoretsky, and N. N. Mikhailov, Phys. Rev. B {\bf 93}, 085308 (2016).
\bibitem {Yakunin2} M. V. Yakunin, S. S. Krishtopenko, S. M. Podgornykh, M. R. Popov, V. N. Neverov, N. N. Mikhailov, S. A. Dvoretsky, JETP Letters {\bf 104}, 403 (2016).
\bibitem {Elena1} Paolo Michetti, Jan C. Budich, Elena G. Novik, and Patrik Recher, Phys. Rev. B {\bf 85}, 125309 (2012).
\bibitem {Pfeffer} Pawe\l{} Pfeffer and W\l{}odek Zawadzki, Phys. Rev. B {\bf 86}, 195303 (2012).
\bibitem {Novik} E. G. Novik, A. Pfeuffer-Jeschke, T. Jungwirth, V. Latussek, C. R. Becker, G. Landwehr, H. Buhmann, and L. W. Molenkamp, Phys. Rev. B {\bf 72}, 035321 (2005).
\bibitem {APJ} A. Pfeuffer-Jeschke, Ph.D. thesis, Physikalisches Institut, Universitt W\"urzburg, Germany, 2000.  
\bibitem {MarIJMP} Micha\l{} Marchewka, Int. J. Mod. Phys. B, http://dx.doi.org/10.1142/S0217979217501375, (2017).
\bibitem {Zyuzin} A. A. Zyuzin, Si Wu, and A. A. Burkov, Phys. Rev. B {\bf 85}, 165110 (2012).
\bibitem {Ominato} Yuya Ominato and Mikito Koshino, Phys. Rev. B {\bf 93}, 245304 (2016).
\bibitem {SolidMar} M. Marchewka, J. Grendysa, D. \.Zak, G. Tomaka, P. \'Sli\.z, E. M. Sheregii, Solid State Comm. {\bf 250}, 104 (2017). 
\bibitem {wrobel} M. Zholudev, F. Teppe, M. Orlita, C. Consejo, J. Torres, N. Dyakonova, M. Czapkiewicz, J. Wr\'obel, G. Grabecki,
N. Mikhailov, S. Dvoretskii, A. Ikonnikov, K. Spirin, V. Aleshkin, V. Gavrilenko, and W. Knap, Phys. Rev. B {\bf 86}, 205420 (2012).
\bibitem {Maier} Luis Maier, Jeroen B. Oostinga, Daniel Knott, Christoph Brüne, Pauli Virtanen, Grigory Tkachov, Ewelina M. Hankiewicz, Charles Gould, Hartmut Buhmann, and Laurens W. Molenkamp, Phys. Rev. Lett. {\bf 109}, 186806 (2012).
\bibitem {boundary} M. G. Burt, J. Phys.: Condens Matter {\bf 11}, R35 (1999); {\bf 4}, 6651 (1992).
\bibitem {ABRIKOSOV1970} A. A. Abrikosov, S. D. Beneslavskii, Soviet Physics JETP {\bf 32}, 699 (1971).
\bibitem {Burkov} A. A. Burkov, Journal of Physics: Condensed Matter {\bf 27}, 113201 (2015).

\end{thebibliography}
\end{document}